\documentclass[preprint,showpacs,preprintnumbers,amsmath,amssymb]{revtex4}

\usepackage{graphicx}
\usepackage{dcolumn}
\usepackage{bm}

\begin{document}


\title{Possible high temperature superconductivity in Ti-doped A-Sc-Fe-As-O (A= Ca, Sr) system}

\author{G. F. Chen$^{1,2}$, T. -L. Xia$^{2}$, H. X. Yang$^{1}$, J. Q. Li$^{1}$, P. Zheng$^{1}$, J. L. Luo$^{1}$, and N. L. Wang$^{1}$}

\affiliation{$^{1}$Beijing National Laboratory for Condensed
Matter Physics, Institute of Physics, Chinese Academy of Sciences,
Beijing 100190, People¡¯s Republic of China}

\affiliation{$^{2}$Department of Physics, Renmin University of
China, Beijing 100872, People's Republic of China}


\begin{abstract}
We report a systematic study on the effect of partial substitution
of Sc$^{3+}$ by Ti$^{4+}$ in Sr$_{2}$ScFeAsO$_{3}$,
Ca$_{2}$ScFeAsO$_{3}$ and Sr$_{3}$Sc$_{2}$Fe$_{2}$As$_{2}$O$_{5}$
on their electrical properties. High level of doping results in an
increased carrier concentration and leads to the appearance of
superconductivity with the onset of T$_{c}$ up to 45 K.

\end{abstract}



\maketitle

Since the discovery of superconductivity in LaFeAsO$_{1-x}$F$_x$
(abbreviated as 1111) with T$_c\sim$ 26 K,\cite{Kamihara08} the
FeAs-based systems have attracted a great deal of research
interest. Substituting La with other rare earth (RE) elements with
smaller ion radii dramatically enhances the T$_c$ up to 41-55
K.\cite{Chen1, Ren1} At room temperature, all these parent
compounds crystallize in a tetragonal ZrCuSiAs-type structure,
which consists of alternate stacking of edge-sharing Fe$_2$As$_2$
tetrahedral layers and RE$_2$O$_2$ tetrahedral layers along
c-axis. Soon after this discovery, another group of compounds
AFe$_2$As$_2$ (A=Ba, Sr, Ca) (122), which crystallize in a
tetragonal ThCr$_2$Si$_2$-type structure with identical
Fe$_2$As$_2$ tetrahedral layers as in LaFeAsO, were also found to
be superconducting with T$_c$ up to 38 K upon hole
doping.\cite{Rotter2, Chen, Canfield} Except for the above
systematically studied series of iron-based superconductors,
attempts have also been tried to find iron-based superconductors
with new structures, resulting in the discovery of superconductors
like A$_{x}$FeAs (A=Li, Na)\cite{Wang,Pitcher,Tapp} and
FeSe$_{1-x}$Te$_{x}$.\cite{Wu}

On the other hand, it seems that critical temperature increases
with increasing the length of separation between the two
neighboring Fe(As,P) layers.\cite{Ivanovskii} Indeed,
Sr$_{2}$ScFePO$_{3}$ (21113) have been reported to be
superconducting with T$_{c}$ up to 17 K, which is the highest in
arsenic-free iron-based oxypnictide systems.\cite{Ogino} The same
structure was previously studied in oxypnictides and
oxychalcogenide systems,\cite{Schafer,Hor,Zhu2} where
superconductivity have been predicted to be existing once the
carries are appropriately introduced into the systems. The crystal
structure of such materials constitutes of one perovskite-like
Sr$_{4}$Sc$_{2}$O$_{6}$ layer alternating with an edge-sharing
Fe$_2$P$_2$ tetrahedral layer along c-axis. Hence the expanded
c-axis lattice constant makes it a good candidate for high
temperature superconductors. In such materials, doping can be
applied at Sr/Sc/O sites within the perovskite-like units and the
Fe$_2$P$_2$ layers as well.
\begin{figure}[h]
\includegraphics[width=15cm,clip]{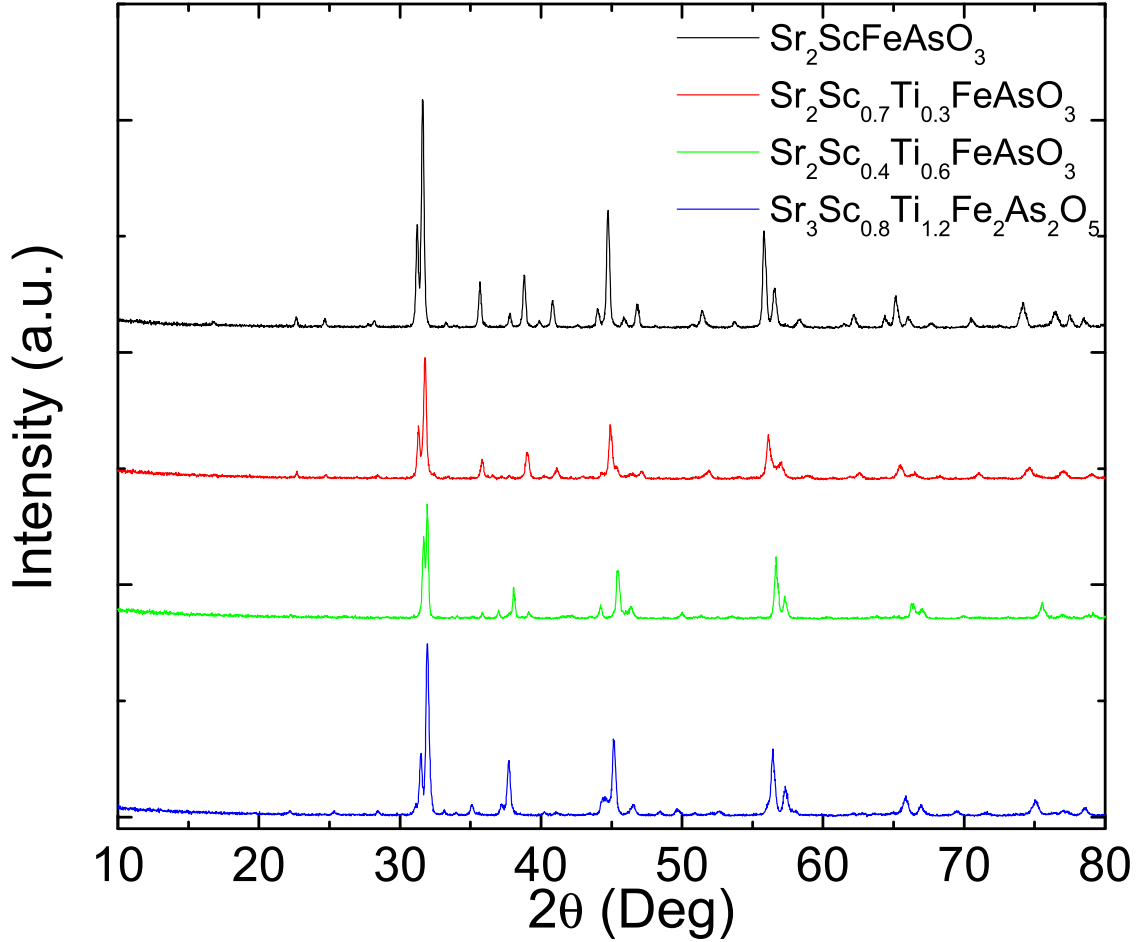}

\caption{(Color online) X-ray powder diffraction patterns for
Sr$_{2}$Sc$_{1-x}$Ti$_{x}$FeAsO$_{3}$ (x=0, 0.3, 0.6) and
Sr$_{3}$Sc$_{0.8}$Ti$_{1.2}$Fe$_{2}$As$_{2}$O$_{5}$.}
\end{figure}
To search for the superconductors with higher critical
temperature, we have also tried different materials with stacking
structure of perovskite-type oxide layer and anti-fluorite
pnictide layer in oxypnictides. In present work, we report on
successful preparation of A$_{2}$ScFeAsO$_{3}$ (A=Ca, Sr) and a
systematic study on effect of partial substitution of Sc$^{3+}$ by
Ti$^{4+}$ in A$_{2}$ScFeAsO$_{3}$ and
Sr$_{3}$Sc$_{2}$Fe$_{2}$As$_{2}$O$_{5}$ (32225) on the electrical
properties. Substituting Sc$^{3+}$ by Ti$^{4+}$ results in an
increased carrier concentration and leads to the appearance of
superconductivity with a possible critical temperature as high as
45 K.

\begin{figure}[h]
\includegraphics[width=15cm,clip]{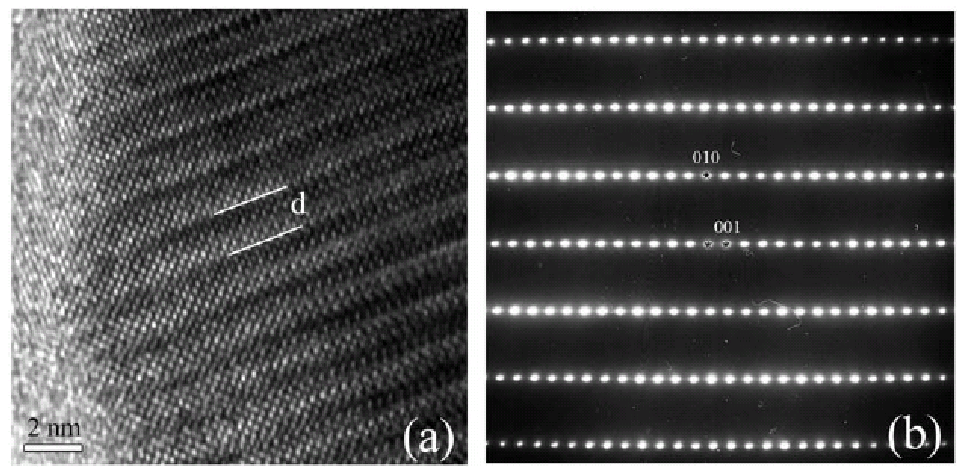}

\caption{(Color online) High-resolution TEM image (a) and
corresponding electron diffraction pattern (b) of a
Sr$_{2}$ScFeAsO$_{3}$ crystal viewed along the [100] zone axis
direction.}
\end{figure}

The polycrystalline samples were prepared by the solid state
reaction method using SrAs, CaAs, SrO, CaO, Sc$_{2}$O$_{3}$,
TiO$_{2}$, Fe, FeAs and Fe$_{2}$As as starting materials. SrAs and
CaAs were obtained by reacting Sr or Ca and As at 800$^{\circ}C$
for 20 h. FeAs and Fe$_{2}$As were prepared by reacting Fe and As
at 900$^{\circ}C$ for 20 h. All the powders were weighed in a
desired ratio and the mixtures were ground thoroughly and
cold-pressed into pellets. The pellets were sealed in quartz tube
under argon atmosphere. They were then annealed for 50 h at a
temperature of 1150$^{\circ}C$. The resulting samples were
characterized by a powder X-ray diffraction(XRD) method with Cu
K$\alpha$ radiation at room temperature. The electrical
resistivity measurements were preformed in a Physical Property
Measurement System(PPMS) of Quantum Design company with standard
4-probe method. The electron diffraction pattern and
high-resolution image were obtained on a FEI Tecnai-F20 (200 kV)
transmission electron microscope (TEM). The TEM samples were
prepared by crushing the polycrystalline parent compound, then the
resultant suspensions were dispersed on a holy carbon-covered Cu
grid.

\begin{figure}[b]
\centerline{\includegraphics[width=15cm]{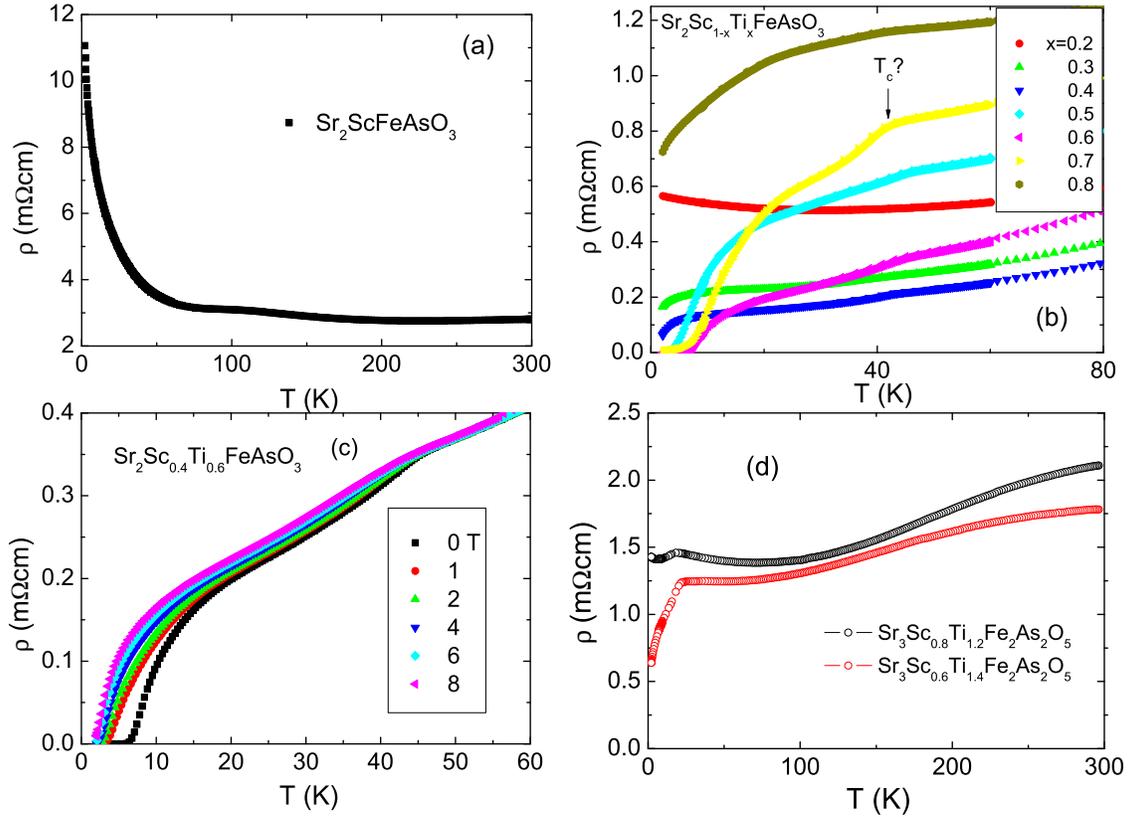}}%
\caption{(Color online) (a)The electrical resistivity vs
temperature for the parent compound Sr$_{2}$ScFeAsO$_{3}$. (b)The
electrical resistivity vs temperature for a series of
Sr$_{2}$Sc$_{1-x}$Ti$_{x}$FeAsO$_{3}$ (x=0.2-0.8). (c)The
electrical resistivity vs temperature for
Sr$_{2}$Sc$_{0.4}$Ti$_{0.6}$FeAsO$_{3}$ in magnetic fields up to 8
T. (d) The electrical resistivity vs temperature for
Sr$_{3}$Sc$_{2-x}$Ti$_{x}$Fe$_{2}$As$_{2}$O$_{5}$ with x= 1.2 and
1.4. }
\end{figure}

XRD patterns for Sr$_{2}$Sc$_{1-x}$Ti$_{x}$FeAsO$_{3}$ (x=0, 0.3,
0.6) and Sr$_{3}$Sc$_{0.8}$Ti$_{1.2}$Fe$_{2}$As$_{2}$O$_{5}$ are
shown in Fig.1. For Sr$_{2}$ScFeAsO$_{3}$ and
Sr$_{2}$Sc$_{0.7}$Ti$_{0.3}$FeAsO$_{3}$, all the main peaks can be
indexed on the basis of tetragonal Sr$_2$FeCuSO$_3$-type structure
with the space group P4/nmmz. The lattice constants for the parent
compound were determined to be a = 4.048 $\AA$ and c = 15.804
$\AA$. However, the high level of doping tends to be more
favorable for the formation of
Sr$_{3}$Sc$_{2-x}$Ti$_{x}$Fe$_{2}$As$_{2}$O$_{5}$ with the space
group of I4/mmm. For comparison, we display also the diffraction
pattern of the sample
Sr$_{3}$Sc$_{0.8}$Ti$_{1.2}$Fe$_{2}$As$_{2}$O$_{5}$, which was
prepared under the same conditions.

Figure 2 shows the typical high-resolution TEM image and
corresponding electron diffraction pattern of the as synthesized
product viewed along the [100] zone axis direction. All the
diffraction spots shown in Fig. 2(b) can be well indexed by the
space group P4/nmm with b $\sim$ 4.0 $\AA$ and c $\sim$ 16 $\AA$,
which coincide well with corresponding values obtained from XRD
patterns. In addition, the high-resolution image taken along the
[100] zone axis shows clear layered structure with the layer
distance d of about 16 $\AA$, which is in good agreement with the
c-axis lattice parameter determined by X-ray diffraction
measurement. As a result, we can confirm that the sample used in
the present TEM study is Sr$_{2}$ScFeAsO$_{3}$ phase. Due to the
different space groups and lattice parameters between 21113 phase
(P4/nmm) and 32225 phase (I4/mmm), they exhibit different electron
diffraction pattern along the [100] zone axis.

\begin{figure}[b]
\centerline{\includegraphics[width=15cm]{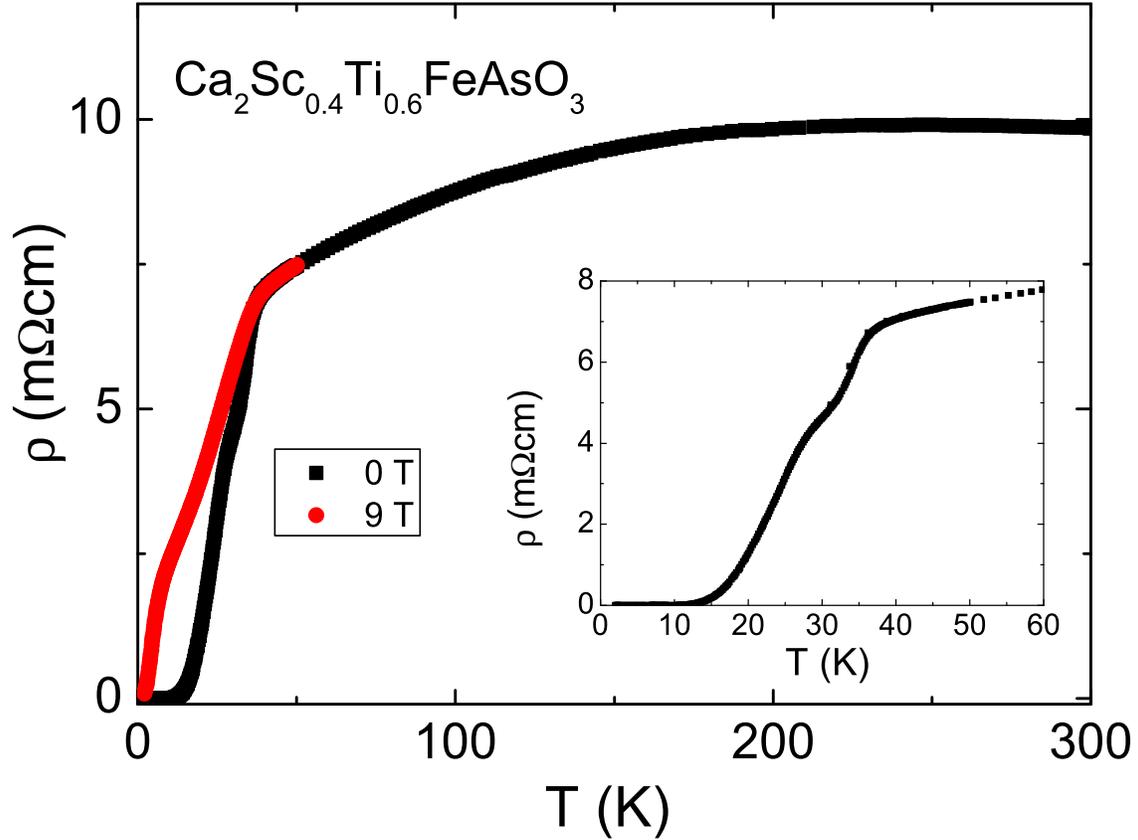}}%
\caption{(Color online) The electrical resistivity vs temperature
for Ca$_{2}$Sc$_{0.4}$Ti$_{0.6}$FeAsO$_{3}$ under 0 T and 9 T. The
inset shows the enlarged part at low temperature under 0 T.}
\end{figure}

Figure 3(a) shows the temperature dependence of the resistivity
for Sr$_{2}$ScFeAsO$_{3}$. The parent compound exhibits a
semiconducting-like behavior. No obvious anomaly is observed in
resistivity. This is in contrast to those of the three prototype
parent compounds with ZrCuSiAs-, ThCr$_2$Si$_2$- and PbFCl-type
structures, in which strong anomalies related to
structural/magnetic phase transitions occur in the parent
compounds. Electron- or hole-doping strongly weakens this anomaly
and induces superconductivity. The existence of competing orders
are the common feature for those compounds and high temperature
superconductivity appears near this instability. Figure 3(b) shows
the temperature dependence of the resistivity for a series of
Sr$_{2}$Sc$_{1-x}$Ti$_{x}$FeAsO$_{3}$ with x=0.2-0.8. With the
Ti$^{4+}$-doping, the resistivity decreases and shows a metallic
behavior. It superconducts when it is doped with 30$\%$ Ti on Sc
sites. While the onset of superconductivity occurs at about 45 K,
full bulk superconductivity is observed until 7 K for the sample
of Sr$_{2}$Sc$_{0.4}$Ti$_{0.6}$FeAsO$_{3}$. The two step-like
phenomena in conductivity may originate from two different
superconducting phases. Figure 3(c) shows the resistivity of
Sr$_{2}$Sc$_{0.4}$Ti$_{0.6}$FeAsO$_{3}$ in magnetic fields up to 8
T. The superconductivity was suppressed slowly by applying the
magnetic fields, indicating the superconductivity is intrinsic. It
is difficult to determine the upper critical fields due to the
broad transition. It is rather interesting that
Sr$_{2}$Sc$_{1-x}$Ti$_{x}$FeAsO$_{3}$ does superconduct even for
the high level of doping with x=0.7. As mentioned above, the high
level of doping possibly leads to the formation of ``32225" phase.
Hence we attempted to dope Sr$_{3}$Sc$_{2}$Fe$_{2}$As$_{2}$O$_{5}$
with the same route. A trace of superconductivity extending up to
near 20 K was observed in
Sr$_{3}$Sc$_{0.8}$Ti$_{1.2}$Fe$_{2}$As$_{2}$O$_{5}$ and
Sr$_{3}$Sc$_{0.6}$Ti$_{1.4}$Fe$_{2}$As$_{2}$O$_{5}$, as shown in
Fig. 3(d). Hence it raises one question: where does the
superconductivity come from, ``21113" phase, ``32225" phase, or
other new phase? Detailed characterization of the superconducting
phase is in progress.

Other members of this family with A= Ca and Ba have also been
prepared and characterized. Indeed, Ti ions doping in
Ca$_{2}$ScFeAsO$_{3}$ can induce superconductivity with the onset
temperature up to 37 K, as shown in Fig. 4. Up to now, no hint of
superconductivity has been found in Ti-doped
Ba$_{2}$ScFeAsO$_{3}$.

In summary, we have succeeded in preparing A$_{2}$ScFeAsO$_{3}$
(A=Ca, Sr) and performed a systematic study on effect of partial
substitution of Sc$^{3+}$ by Ti$^{4+}$ in A$_{2}$ScFeAsO$_{3}$ and
Sr$_{3}$Sc$_{2}$Fe$_{2}$As$_{2}$O$_{5}$ on the electrical
properties. Superconductivity with the onset of T$_{c}$ up to 45 K
could be induced by high level electron doping. Our results
indicate that it is possible to find high temperature
superconductivity in other layered oxypnictides.

\begin{acknowledgments}
This work is supported by the NSFC, CAS, and the 973 project of
the MOST of China.

\end{acknowledgments}

\end{document}